\documentclass{aa}

\usepackage{natbib,twoopt}
\usepackage[breaklinks=true]{hyperref} 
\bibpunct{(}{)}{;}{a}{}{,} 
\makeatletter
\newcommandtwoopt{\citeads}[3][][]{\href{http://adsabs.harvard.edu/abs/#3}%
{\def\hyper@linkstart##1##2{}%
\let\hyper@linkend\@empty\citealp[#1][#2]{#3}}}
\newcommandtwoopt{\citepads}[3][][]{\href{http://adsabs.harvard.edu/abs/#3}%
{\def\hyper@linkstart##1##2{}%
\let\hyper@linkend\@empty\citep[#1][#2]{#3}}}
\newcommandtwoopt{\citetads}[3][][]{\href{http://adsabs.harvard.edu/abs/#3}%
{\def\hyper@linkstart##1##2{}%
\let\hyper@linkend\@empty\citet[#1][#2]{#3}}}
\newcommandtwoopt{\citeyearads}[3][][]%
{\href{http://adsabs.harvard.edu/abs/#3}
{\def\hyper@linkstart##1##2{}%
\let\hyper@linkend\@empty\citeyear[#1][#2]{#3}}}
\makeatother

\usepackage{graphicx}
\usepackage{txfonts}
\usepackage{natbib}
\usepackage{color} 
\usepackage{verbatim}

\begin{document}

\title{Charging of small grains in a space plasma:\\
Application to Jovian stream particles}

   \author{A. R. Dzhanoev
          \inst{1},
          J. Schmidt\inst{2},
          X. Liu\inst{2},
          \and F. Spahn \inst{1}
          }

   \institute{Universit\"at Potsdam, Karl-Liebknecht-Str. 24/25 Building 28, 14476 Potsdam-Golm, Germany\\
              \email{dzhanoev@uni-potsdam.de}
         \and
             University of Oulu, Astronomy and Space Physics, PL 3000, Oulu, Finland\\
             }

   \date{Received .., 2015; accepted ...}

\titlerunning{Charging of small grains in a space plasma}
\authorrunning{A. R. Dzhanoev, J. Schmidt et al.,}
                                
\abstract  
{Most theoretical investigations of dust charging processes in space have treated the current balance condition as independent of grain size. 
However, for small grains, as they are often observed in space environments, a dependence on grain size is expected due to secondary electron emission (SEE). 
Here, by the term "small" we mean a particle size comparable to the typical penetration depth for given primary electron energy (Draine \& Salpeter \citeyear{DraineSalpeter1979}, Chow et al. \citeyear{Chowetal1993}). The results are relevant for the dynamics of small, charged dust particles emitted by the volcanic moon Io, which form the Jovian dust streams.}
{We revise the theory of charging of small (sub-micron sized) micrometeoroids to take into account a high production of secondary electrons for small grains immersed in an isotropic flux of electrons. We apply our model to obtain an improved estimate for the charge of the dust streams leaving the Jovian system, detected by several spacecraft.}
{We apply a continuum model to describe the penetration of primary electrons in a grain and the emission of secondary electrons along the path. Averaging over an isotropic flux of primaries, we derive a new expression for the secondary electron yield, which can be used to express the secondary electron current on a grain.}
{For the Jupiter plasma environment we derive the surface potential of grains composed of NaCl (believed to be the major constituent of Jovian dust stream particles) or silicates. For small particles the potential depends on grain size and the secondary electron current induces a sensitivity to material properties. As a result of the small particle effect, the estimates for the charging times and for the fractional charge fluctuations of NaCl grains obtained using our general approach to SEE give results qualitatively different from the analogous estimates derived from the traditional approach to SEE. We find that for the charging environment considered in this paper field emission does not limit the charging of NaCl grains.}
{}


\maketitle

\section{Introduction}
A considerable number of mechanisms can contribute to the charging of dust grains in space.
Generally, the charging process of a cosmic grain can be described by the continuity equation in integral form
\begin{equation}
\label{charging}
dq_d/dt+\sum_kJ_k(\phi_d(t))=0,
\end{equation}
where the $J_k$ represent relevant charging currents.  For spherical particles, the relation between charge $q_d$ and surface potential $\phi_d$ is given by $q_d=4\pi\varepsilon_0 a\phi_d$ provided that the grain radius $a$ is much smaller than the plasma Debye length $\lambda_d$ and the collisional mean free path of the plasma particles, $\lambda$, is generally larger than $\lambda_d$. Here $\epsilon_0=8.854\cdot10^{-12}\text{ F}\cdot\text{m}^{-1}$ is the vacuum permittivity. The currents $J_k$ are functions of the grain properties, the ambient plasma conditions and the radiation field. At equilibrium the net current to the grain vanishes and the equilibrium grain potential $\phi_d$ is obtained from the flux balance equation $\sum_kJ_k(\phi_d(t))=0$. It is recognized that the secondary electron emission yield (the ratio of emitted electrons to incident ones) might be particularly important in charging of small cosmic grains. In space, attention has focused on SEE from spacecraft surfaces or small dust particles caused by hot electrons in planetary magnetospheres (Draine \& Salpeter \citeyear{DraineSalpeter1979}, Whipple \citeyear{Whipple1981}, Hor{\'a}nyi \citeyear{Horanyi92}, Kempf et al. \citeyear{Kempf}). The conventional (Whipple \citeyear{Whipple1981}, Meyer-Vernet \citeyear{MeyerVernet}, Hor{\'a}nyi \citeyear{Horanyi96}) model for the secondary electron current $J_{s}$ on grains immersed in a plasma uses the approach suggested by \citet{Sternglass1957} to estimate SEE. However, this approach is valid only for the case of big grains (semi-infinite slab model).
\citet{DraineSalpeter1979} have shown that the secondary electron yield is enhanced when the dimensions of the grains are comparable to the primary electron penetration depth, the so-called small particle effect. \citet{Chowetal1993} have extended the theory suggested by \citet{Jonker1952} to treat the small particle effect of electron impact on small spherical grains. Their model makes the simplifying but unrealistic assumption that the primary electrons are incident normally to the grain surface. 
For grains in a space plasma an isotropic incidence is more realistic. Namely, in typical magnetospheric environments in the solar system, 
the relative motion of a dust particle (on a Keplerian orbit) and the nearly co-rotating plasma is small, compared to the thermal velocity of the electrons. Therefore, the flux of electrons can be treated as quasi isotropic. First attempts to include the assumption of isotropic incidence into the model of SEE 
from spherical grains were made by \citet{Chow1997}, and \citet{KimuraMann}. In this paper, we reconsider the formulation of a model for the secondary electron current $J_{s}$ on grains immersed in an isotropic flux of electrons, basing on the results by \citet{Dzhanoev2015} that give a better match of experimental data than previous SEE models. In this approach we use a Maxwellian distribution of electron velocities. We illustrate our results by applying them to dust charging precesses in the Jupiter environment. Specifically, the small size effect of SEE will be important for the Jovian small particles, i.e. dust particles with radii in the range of $5\text{ nm}\leq a\leq15\text{ nm}$ (for grains with speed $\varv >200 \text{ km s}^{-1}$, Zook et al., \citeyear{Zook1996}) and $30\text{ nm}\leq a \leq0.3$ $ \mu\text{m}$ (for grains with speed $20\text{ km s}^{-1}\leq \varv \leq 56\text{ km s}^{-1}$, Gr\"un et al. \citeyear{Gruen1996}) emanating from the Jovian system.
 
\section{Method and model}
\label{second}
\subsection{Generalization of the SEE model}
\label{third}
Typically, secondary electron emission depends on the material, being small in metals, but often exceeding the incident flux in insulators. It has been suggested that the energy dependence of the SEE yield can be described by the \citet{Sternglass1957} universal curve, when the yield is normalized by the maximum yield and the primary electron energy by the energy where the yield is maximized. However, a series of measurements of SEE covering a wide range of primary energies (Salow \citeyear{Salow},Young \citeyear{Young}) shows that the theory of Sternglass fails to fit the experimental data at high primary electron  energy. The empirical formula developed by \citet{DraineSalpeter1979} generally shows a better agreement with experiments but overestimates the experimental data. Later, \citet{Chowetal1993} have modified the yield equation by \citet{Jonker1952} and derived the yield for secondary emission from a spherical dust grain (also for grain sizes that are comparable to the penetration depth of the primary electrons). However, their treatment was limited to the case of primary electrons that are incident normally to the grain surface. In contrast, a plasma environment in space typically provides an isotropic flux of electrons.

To describe SEE due to isotropic incidence of primary electrons $e_p^-$ of energy $E_0$ on spherical objects of radius $a$ we consider the expression (Dzhanoev et al. \citeyear{Dzhanoev2015}) for the secondary electron yield
\begin{eqnarray}
\label{IsoIncid}
\delta(E_0,a,\theta)=\int_0^{R(E_0)}{\frac{1}{\epsilon} \left(-\frac{dE}{dx} \right)H\left(2a\cos\theta-x\right)dx} \times\\
\nonumber \times \int_0^{\varphi_c}{\exp\left({-\frac{l}{\lambda}}\right)\sin \varphi \displaystyle \,d \varphi}.
\end{eqnarray}
For the evaluation of the yield we must average over the angle of incidence $\theta$, as
\begin{eqnarray}
\label{IsoAngle}
\left<\delta(E_0,a)\right>_{\theta}=\int_0^{\theta_c}{\delta(E_0,a,\theta) \sin \theta \cos \theta \,d \theta}.
\end{eqnarray}
Here, following \citet{Jonker1952} we consider the case when secondary electrons are generated at a distance $x$ from the entry location of the primary electron into a spherical grain and move to the target's surface at an angle $\varphi$ with direction to the nearest surface point. The Heaviside function $H(2a\cos\theta-x)$ selects contributions to SEE only from primaries traveling on a straight path lying entirely in the spherical grain. The parameter $\lambda$ denotes the mean free path
\begin{equation}
\label{lambdaIso}
\lambda=(An)^{-1}(E_m/\bar{r}_m)^n=R(E_m)/\bar{r}_m^n
\end{equation}
and $\epsilon$ is the energy necessary to produce one secondary electron
\begin{equation}
\label{epsilonIso}
\epsilon=(E_m/\delta_m)[\hat{G}_n(\bar{r}_m)/\bar{r}_m],
\end{equation}
where where $\delta_m$ is the maximum yield and $E_m$ is the corresponding energy, and (Dzhanoev et al. \citeyear{Dzhanoev2015})
\begin{eqnarray}
\label{Giso}
\hat{G}_n(\bar{r}) \equiv\int_0^1{\mu^{1-\frac{1}{n}} d\mu}\int_0^1{z^{\frac{1}{n}}\exp\left(-\mu\frac{\bar{r}^n}{z}\right) dz} \times \\
\nonumber \times \int_0^{\mu(\bar{r}/z^{\frac{1}{n}})}{\exp(y^n)\, dy}.
\end{eqnarray}
The function $\hat{G}_n(\bar{r})$ has a single maximum at $r=\bar{r}_m$. 
$R(E_0)=(An)^{-1}E_{0}^{n}$ is the so-called projected range. The parameters $A$ and $n$ depend on the projectile and the target material and are determined by experimental measurement of $R(E_0)$, giving $n=1.5$ for electrons (Fitting \citeyear{Fitting}). 
The distance $l$ to the grain surface is given by $l=\sqrt{a^2-(\varw a)^2\sin^2\varphi}-\varw a\cos\varphi$ with $\varw=\sqrt{1+x^2/a^2-2(x/a)\cos \theta}$.
We note that for a spherical dust grain, $\varphi$ can vary from $0$ to $\displaystyle\varphi_c=\displaystyle\pi$.
The angle $\theta$  between the incident direction of the primary and the normal to the surface of the grain can vary from $0$ to 
$\theta_c=\displaystyle\frac{\pi}{2}$. Note that when dealing with a grain of radius $a$ one can obtain from equation (\ref{IsoAngle}) an expression for the secondary electron yield valid for a semi-infinite slab by taking the limit $x/a\rightarrow0$. 

\subsection{Generalized model for the secondary electron current}
\label{forth}
In what follows, we obtain a new expression for the secondary electron current $J_{s} (\phi_d,a)$ depending on grain radius, which includes the case of small grains. This expression generalizes existing expressions in the literature (Meyer-Vernet \citeyear{MeyerVernet}, Chow et al. \citeyear{Chowetal1993}). For a Maxwellian distribution of primary electron velocities, with the help of equation (\ref{IsoAngle}) we obtain
\begin{equation}
\label{extsecurrent}
J_{s} (\phi_d,a)= \delta_m J_{e}^0e^{-\chi_e}\left\{
\begin{array}{l l}
\displaystyle F_{0}(E_m/kT_e,a) &  \text{$\chi_{e}>0$}\\
F_{b}(E_m/kT_e,a)(1-\chi_s)e^{\chi_{s}} & \text{$\chi_{e}\leq0$}
\end{array} \right.
\end{equation}
Here $J^0_{e}=\pm e\pi a^2n_{e}\bar{\varv}_{e}$ and $n_{e}$ and $\bar{\varv}_{e}$ are the number density and the mean thermal speed of the electrons.
Moreover,
\begin{equation}
\label{F4}
F_{h}(x,a)=x^2\int_h^{\infty} u\,\hat{\delta}(uE_m,a)\,e^{-ux} du,
\end{equation} 
where $\hat{\delta}$ is the SEE yield (\ref{IsoAngle}) normalized by $\delta_m$. Further, $b=-\chi_e/(E_m/kT_e)$, $\chi_{e,s}=- e\phi_d/kT_{e,s}$, and $kT_s$ ($1-5\text{ eV}$) - the mean energy of the secondary electrons if their velocity distribution is also approximated by a Maxwellian distribution (Prokopenko \& Laframboise \citeyear{ProkopenkoLaframboise1980}, Goertz  \citeyear{Goertz1989}, Jurac et al. \citeyear{Jurac1995}). We assume that the charging currents are calculated by considering the Orbital-Motion Limited (OML) theory (Laframboise \& Parker \citeyear{LaframboiseParker1973}). In our study, we neglect the secondary emission due to ion impact, and the reflection and backscattering of electrons and ions. In the low energy regime the secondary electron flux cannot be approximated by a Maxwellian independently of the primary electron energy. The assumption of a Maxwell distribution for secondary electrons implies that for small incident energies a significant portion of the secondaries escape with energies greater than the incident energy. Thus, the charging calculation needs to be modified, when the electron energy is tens of eV or less (Goertz  \citeyear{Goertz1989}, Jurac et al. \citeyear{Jurac1995}). However, in Section \ref{SEE}, we show that the small particle effect is negligible for low energy primaries. So, for the purposes of our study the modification of the charging scheme is not necessary.

\section{Results and Discussion}
\label{numerics}
The present model (\ref{extsecurrent}) for the secondary electron current on spherical grains has a wide range of applicability. Here, we focus on the estimation of the electric surface potential of small dust particles in the Jupiter environment.

\subsection{Plasma environment}
\label{subsect3_1}
We use the DG83 plasma model (Divine \& Garrett \citeyear{divine1983charged}) which describes the plasma (electrons, protons and heavy ions) distributions in the Jovian magnetosphere. This model with minor modifications was used by \citet{garrett2008modeling}, and \citet{garrett2012jovian}, and will also be used here. In the following, a summary of the DG83 model is presented, which is mainly based on \citet{divine1983charged}. 
Three plasma populations are included in this model: radiation belt electrons and protons, warm (intermediate energy) electrons and protons, and a cold plasma population consisting of electrons, protons and typical ions found in the Jupiter system ($\mathrm{O}^+$, $\mathrm{O}^{++}$, $\text{S}^+$, $\mathrm{S}^{++}$, $\mathrm{S}^{+++}$, $\mathrm{Na}^+$). 
The temperatures for warm protons and warm electrons are $kT_p^{warm}=30\text{ keV}$, and $kT_e^{warm}=1\text { keV}$, respectively. The temperature for cold protons and ions is the same as for cold electrons and depends on distance from Jupiter (see Section \ref{Secthree}).

The radiation belt particles with high energy will pierce through small grains without interaction and therefore will not contribute to the charging process (Hor{\'a}nyi and Juh{\'a}sz \citeyear{horanyi2010plasma}). Thus, only the other two populations are considered for the charging of small grains. The DG83 model is fully three-dimensional, and the plasma distribution is described in the left-handed Jupiter System III (1965) (Seidelmann \& Divine \citeyear{SeidelmannDivine}). We must also mention that we will use this model in the orbital range $[4R_J, 8R_J]$. In this paper, we evaluate the dust charging at fixed 110$^\circ$ western longitude and 0$^\circ$ latitude, because the plasma distribution is approximately symmetric with respect to 110$^\circ$ western longitude and the Jovian equator according to the formulae in \citet{divine1983charged}. We consider these plasma conditions as typical for the Jupiter environment. Two charge neutrality conditions are assumed for the DG83 model: first, the sum of the number density of cold protons and warm protons is equal to the number density of the warm electrons; second, the total charge carried by cold ions is equal to the that carried by the cold electrons. To get an overall impression of the warm and cold plasma parameters (number density and temperature) used in our charging calculation, the reader is referred to Figs.~6 and 10 in \citet{divine1983charged}.

\subsection{Secondary electron emission} 
\label{SEE}
The dominant source of the Jovian dust streams is likely the moon Io (located at $r=5.9R_J$, where $R_J=71,492\text{ km}$ -- Jupiter radius) and its volcanic activity (Graps et. al \citeyear{Graps2000}). The Jovian stream particles are mainly composed of NaCl and sulphurous compounds, while silicates may be present as a minor constituent (Postberg et al., \citeyear{Postberg2006}). In what follows, we evaluate our grain charging model for NaCl particles, using $\delta_m=6$ and $E_m=600\text{ eV}$ (Bruining, \citeyear{Bruining1954}).
The bulk density for NaCl is given by $\rho=2.2 \times 10^3$ $\text{kg}/\text{m}^3$. We also give estimate of the grain equilibrium potential for silicate particles ($\delta_m=2.4$, $E_m =400\text{ eV}$, (Kollath, \citeyear{Kollath}),  $\rho=3.3 \times 10^3$ $\text{kg}/\text{m}^3$). The thermal energies of primary electrons are distributed according to a Maxwellian distribution.

\begin{figure}[ht!]
{\includegraphics[scale=0.3]{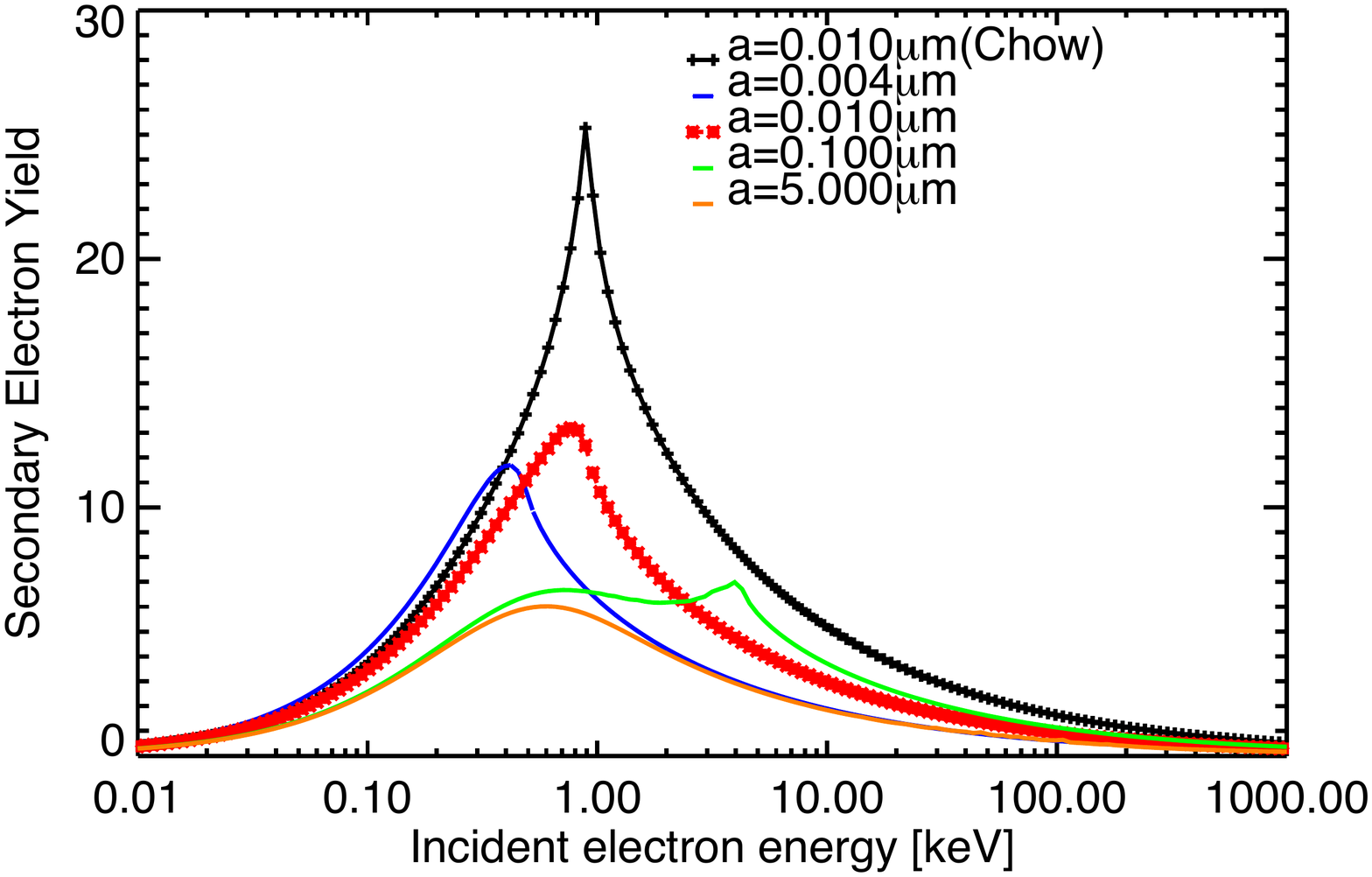}} 
{\includegraphics[scale=0.3]{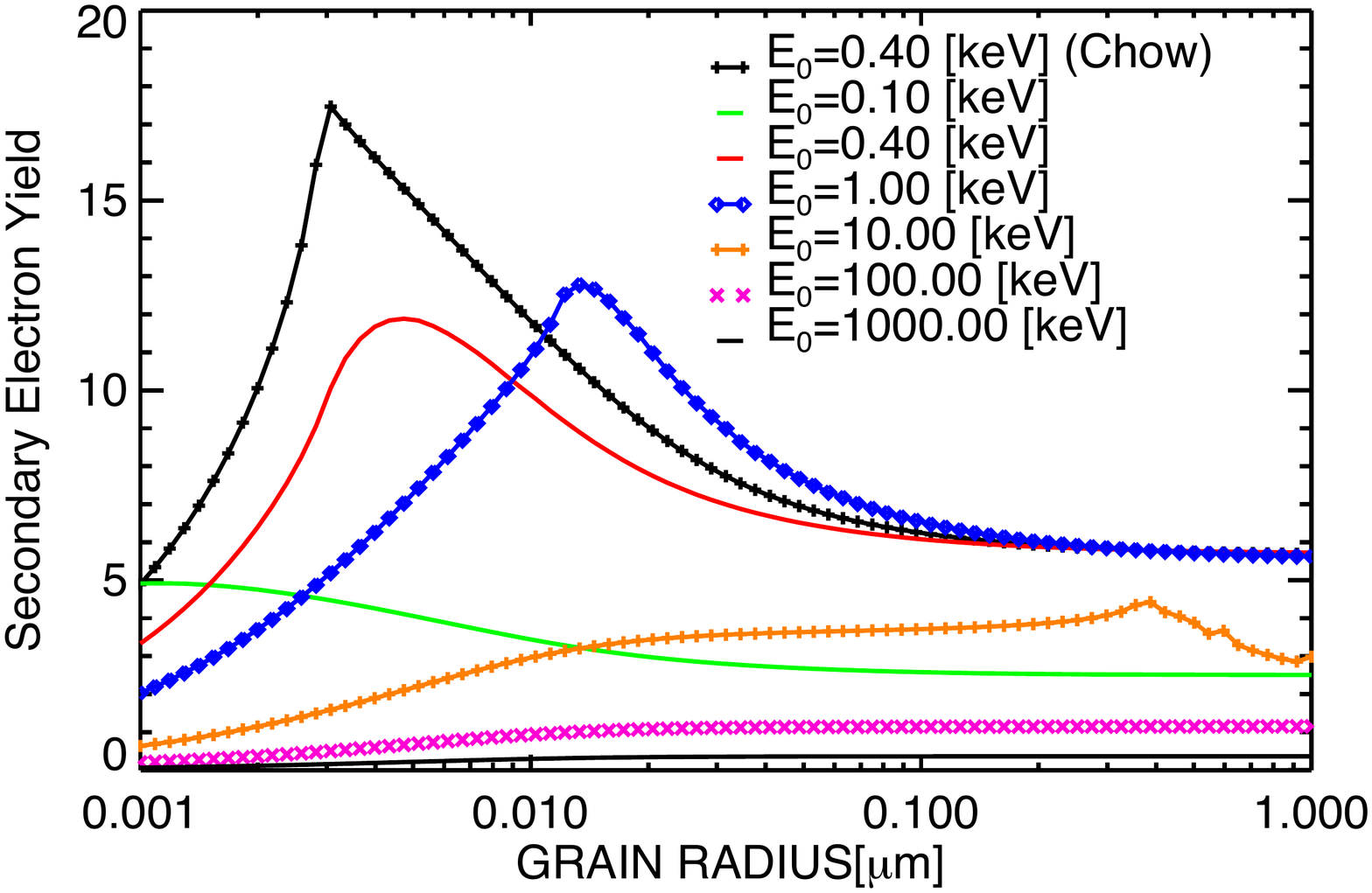}}
\caption{ (Color online). (Top) SEE yield from a NaCl grain for isotropic incidence of electrons, using typical plasma parameters for the Jovian system (see text). Grains of different radii are considered: (i) $a=0.004$ $\mu \text{m}$, (ii) $a=0.01$ $\mu \text{m}$, (iii) $a=0.1$ $\mu \text{m}$ and (iv) $5\mu\text{m}$, the latter corresponding to the limit of big radius $x/a\rightarrow0$; (Bottom) Dependence of SEE yield from NaCl grains on the grain radius for various primary electron energies. On both figures the SEE yield $\delta(E_0,a,\theta=0)$ induced by electrons normally incident to the grain surface is shown for comparison (Chow et al. \citeyear{Chowetal1993}).}
\label{Yields}
\end{figure}
Results for the case of a spherical NaCl grain are illustrated in  Fig.~\ref{Yields}. The top panel shows the secondary electron yield from single particles of different radii. 

Grains with radius $a=0.1\mu\text{m}$ and smaller exhibit a higher yield than larger grains. In the large grain limit the SEE yield converges to that from a flat surface, also revealing the universal energy dependence. The highest value for secondary electron yield is obtained when $R(E_m)\approx a$. This implies that at certain values of the penetration depth $R(E_0)$ the production of secondaries is maximized. 
The bottom panel shows the influence of the primary electron energy on the SEE yield depending on the size of the grain. 
The effect of small grains becomes more pronounced as the electron energy increases from approximately $100\text{ eV}$ to a few $\text{keV}$ (intermediate electron energies).  
For small ($a<0.1\mu\text{m}$) grains the SEE yield does not reveal the universal dependence on energy as it is observed for large ($a>1\mu\text{m}$) grains and flat surfaces. 
\begin{figure}[ht!]
{\includegraphics[scale=0.3]{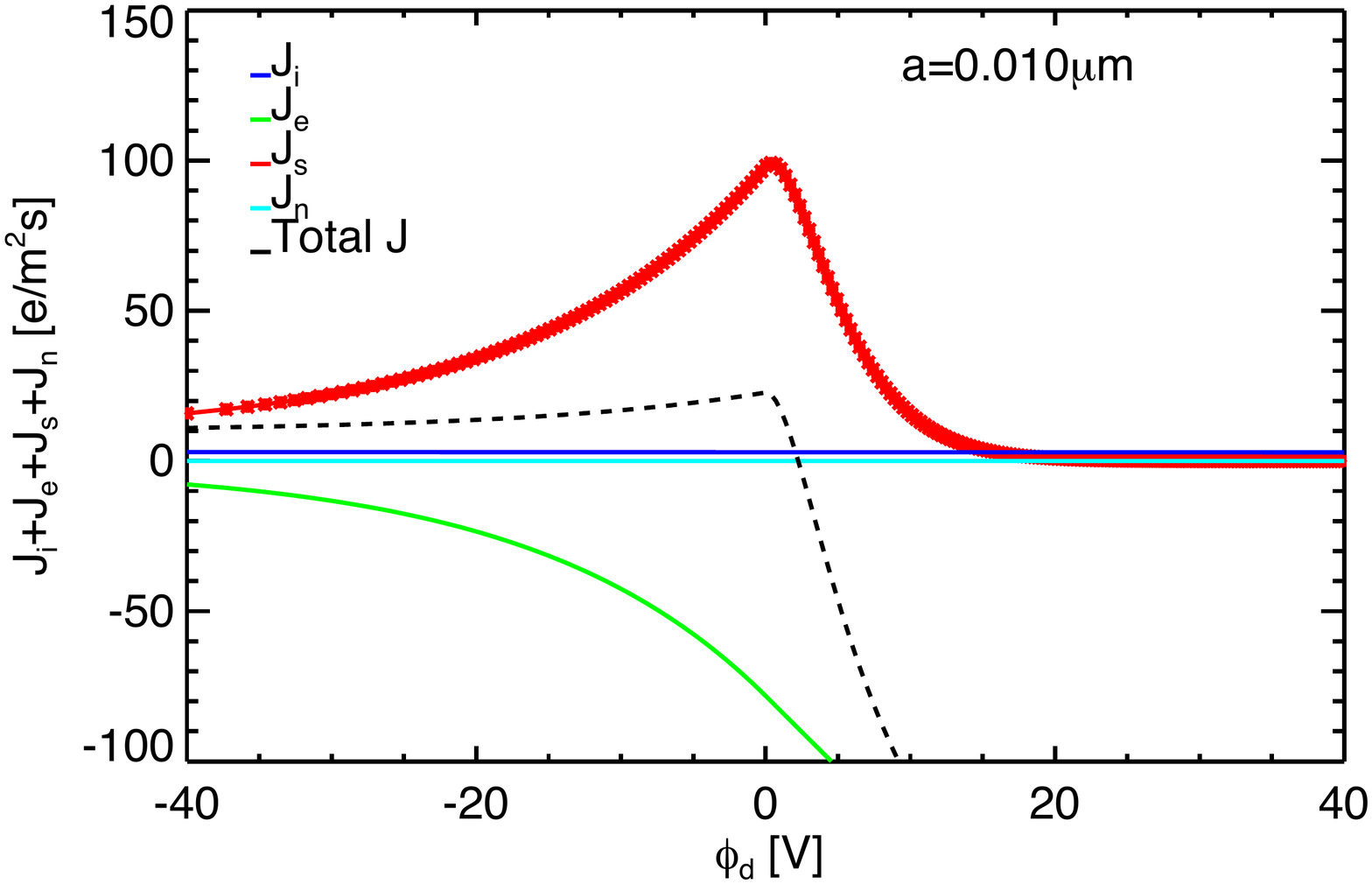}}
{\includegraphics[scale=0.3]{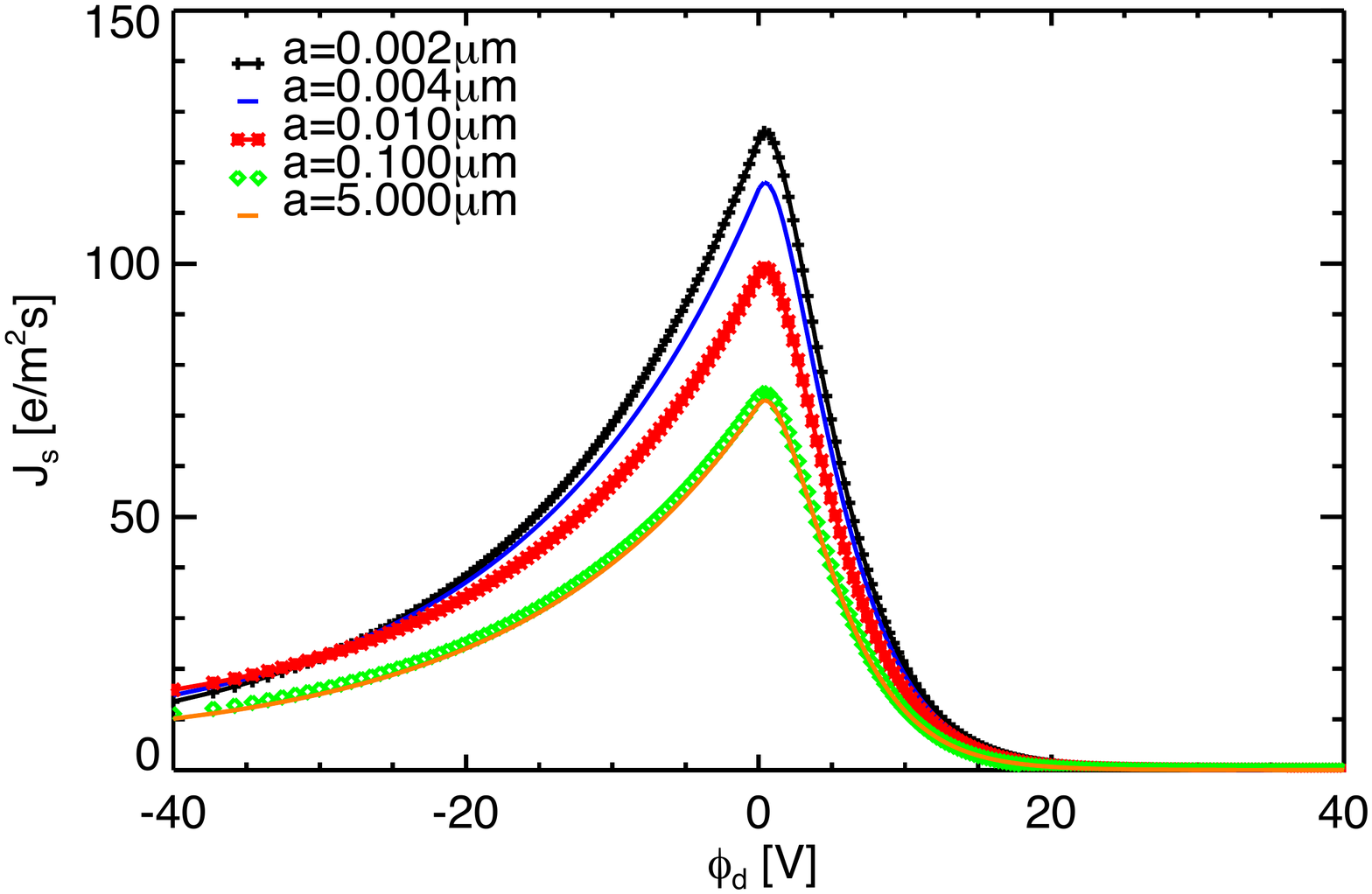}}
\caption{ (Color online). (Top) The current-potential characteristics for NaCl grains with $a=0.01\mu\text {m}$ in the Jupiter plasma environment at $r\sim5.7R_J$ (in the Jupiter equator and at fixed 110$^\circ$ western longitude, see text) with $kT_{e}^{\text{warm}}=1\text{ keV}$, $kT_{e}^{\text{cold}}=0.016\text{ keV}$, and $kT_s=2.5\text{ eV}$. All currents contributing to charging process are shown; (Bottom) Dependence of the secondary electron current $J_{s}(\phi_d,a)$ given by (\ref{extsecurrent}) on the potential of NaCl grains of different radii corresponding to Fig.~\ref{Yields} (top).}
\label{CurrentsSGE}
\end{figure}
Thus, SEE from a spherical grain of given radius with isotropic incident flux converges in the limit of big grain radius to the case of SEE from a semi-infinite slab model. This convergence of $\left<\delta^{grain}(E_0,a)\right>_{\theta}$ to $\delta^{slab}(E_0,\infty,\theta=0)$ is seen in the convergence of all curves to the case of $5\mu\text{m}$ particles, which represents practically the large grain limit. Note also that there is a second peak appearing in the yield curves (Fig.~\ref{Yields}, $a=0.1\mu\text{m}$), which represents an effect due to the finite size of grains. More detailed analysis on this topic can be found in \citet{Dzhanoev2015} and references therein. For comparison, we also show in 
Fig.~\ref{Yields} (top) and (bottom) the results for the SEE yield for the case of normally incident electrons (Chow et al. \citeyear{Chowetal1993}). 
We find that at intermediate electron energies, in the case of isotropic incidence, an electron hitting the grain obliquely is more likely to traverse the small ($a\lesssim0.1\mu\text{m}$) grain. This reduces the SEE yield, since most secondary electrons are produced near the end of an electron path.

\begin{figure}[ht]
\includegraphics[scale=0.3]{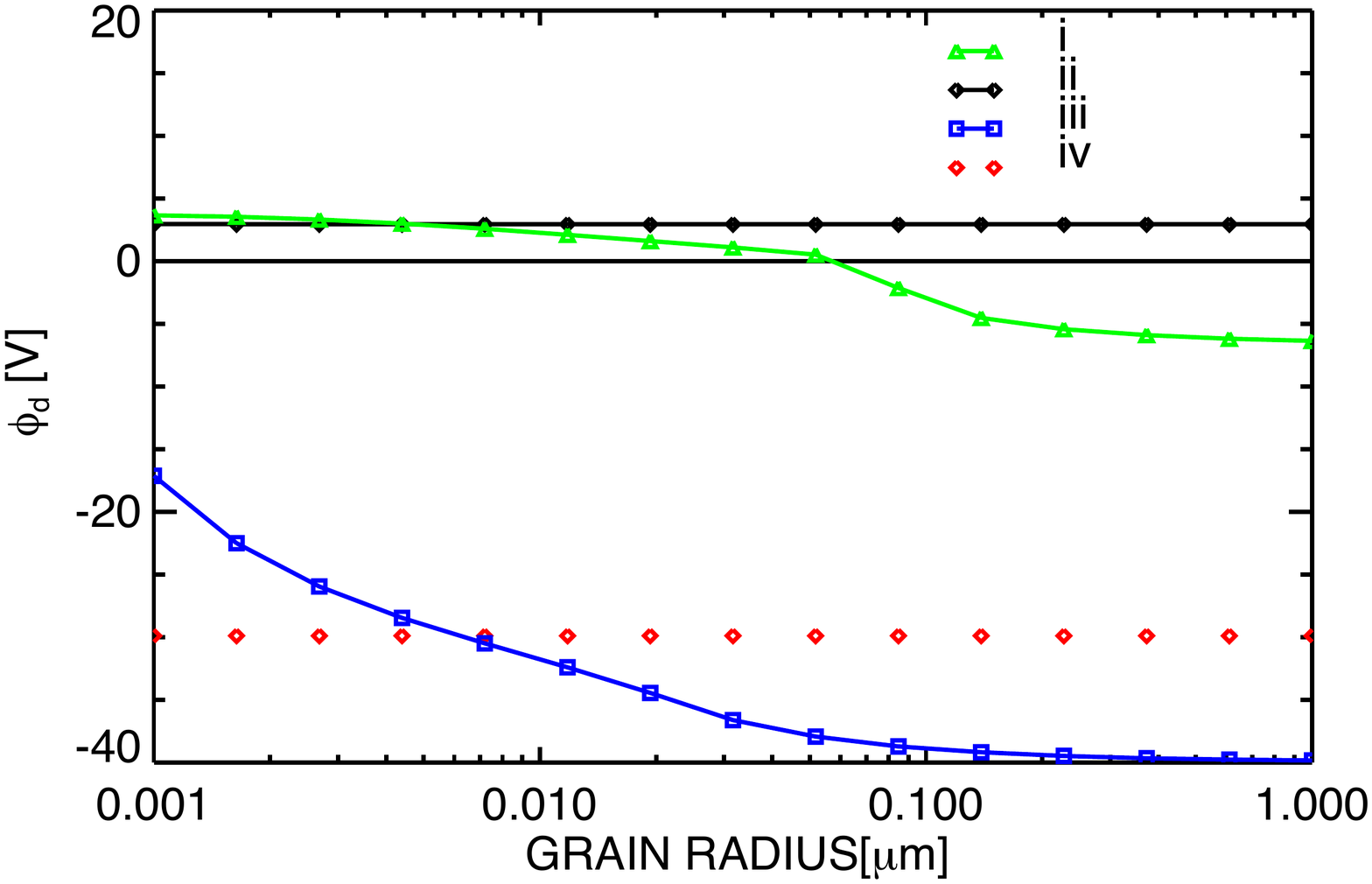}
\includegraphics[scale=0.3]{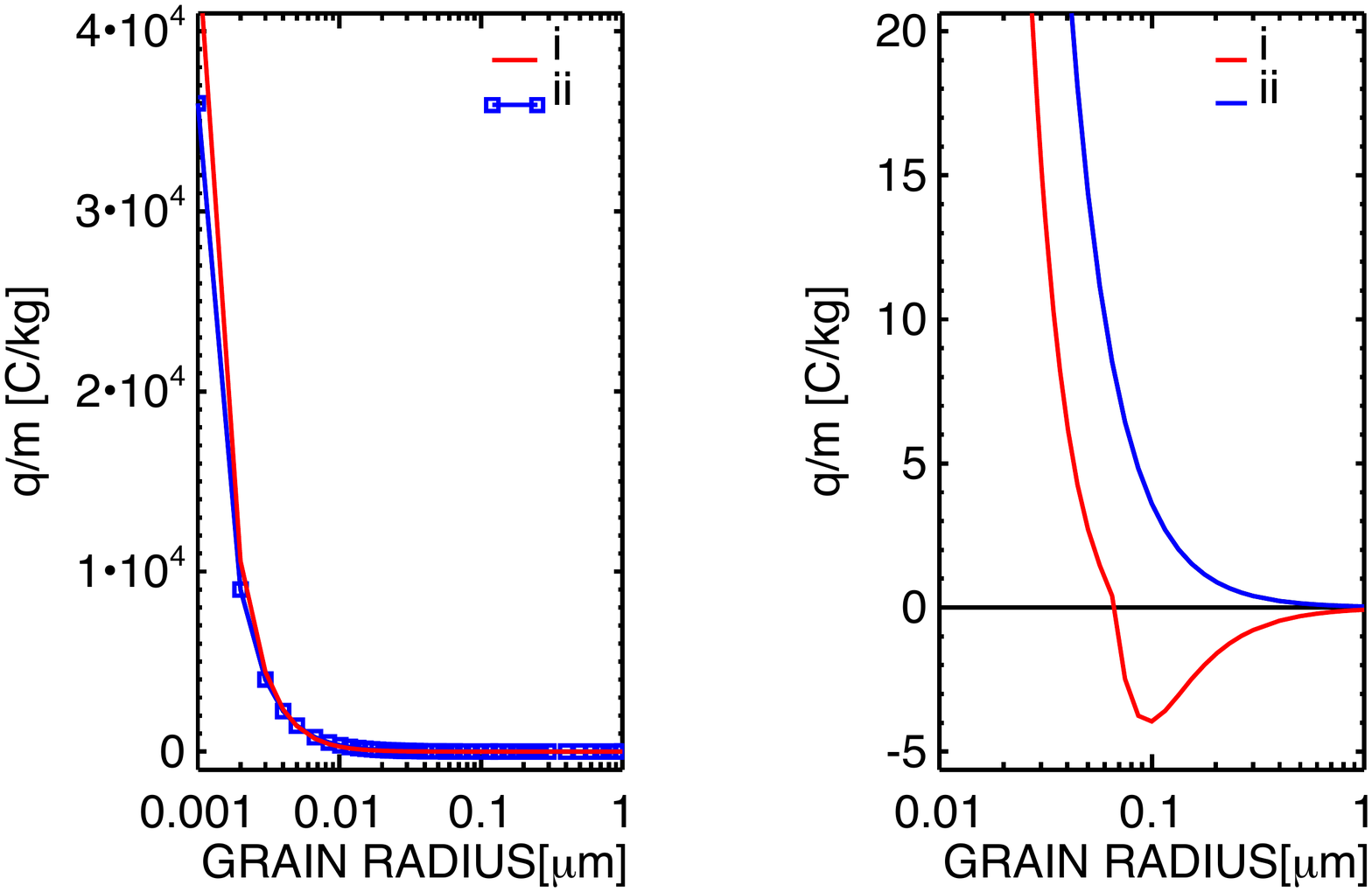}
\caption{(Color online). (Top) Dependence of the equilibrium surface potential $\phi_d(a)$ on grain radius $a$. Materials are NaCl (i), (ii) and silicate (iii), (iv) immersed in typical plasma conditions expected at $r\sim5.7R_J$ distance in the Jupiter system (see text). The results (i) and (iii) are obtained from the generalized $J_s$ model (\ref{extsecurrent}) and compared to results (ii) and (iv) using the Sternglass approximation for SEE $\delta(E_0)=7.4\delta_m(E_0/E_m)\exp(-2\sqrt{E_0/E_m})$; (Bottom, Left) Dependence of the charge-to-mass ratio $q/m$ on grain radius $a$ for NaCl grains corresponding to the results (i) and (ii) in Fig.~\ref{Potential} (top);  
(Bottom, Right) Zoom for small $|q/m|$ emphasizing the existence of a size range where the new SEE model leads to a different polarity of the grains.} 
\label{Potential}
\end{figure} 
The approximation of the secondary electron yield by a semi-infinite slab model may work well for grains of radii $a\geq1\mu\text{m}$ but it is not appropriate for  small grains with sizes comparable or smaller than the mean depth of production of secondary electrons.
So, for particles with radii from nanometers to submicrometers released from Io, the SEE yield increases with the thermal energy of plasma primary electrons $kT_e$ for low $kT_e<E_m$, it dominates when $kT_e$ is of the order of $E_m$, and it decreases when $kT_e\gg E_m$.

\subsection{Secondary electron current and equilibrium potential}
\label{Secthree}
The total current versus surface potential for NaCl grains in the Jupiter plasma environment, evaluated at a distance close to Io (in the equatorial plane at 110 deg western longitude, see Section \ref{subsect3_1}) is shown in Fig.~\ref{CurrentsSGE} (top). Again, we used Eq. (\ref{IsoAngle}) to calculate SEE for an isotropic electron flux with Maxwellian velocity distribution. In this figure we also show all relevant currents generated by plasma electrons ($J_e$) and ions ($J_i$), secondary electron production ($J_s$) calculated via (\ref{extsecurrent}), and the photoelectric effect ($J_n$). As it was shown in Fig.~\ref{Yields} in the limit of big grain radius ($x/a\rightarrow0$), the yield $\left<\delta^{grain}(E_0,a)\right>_{\theta}$ converges to $\delta^{slab}(E_0,\infty,\theta=0)$. 
\begin{figure}[ht]
\includegraphics[scale=0.3]{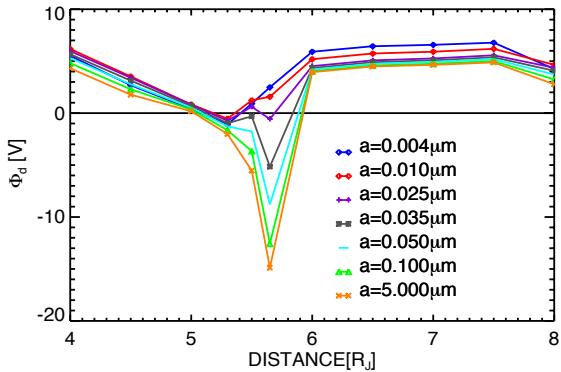}
\caption{(Color online) Equilibrium potential of NaCl grains of different radii as a function of distance from Jupiter.}
\label{BigPotential}
\end{figure}
Correspondingly, the general expression (\ref{extsecurrent}) for $J_{s}(\phi_d,a)$ in the big grain limit converges to $J_{s} (\phi_d,\infty)$, calculated using a semi-infinite slab approximation for SEE, becoming independent of grain size. 
This convergence is shown in Fig.~\ref{CurrentsSGE} (bottom).

In Fig.~\ref{Potential} (top) we present results for the equilibrium potential $\phi_d(a)$ of NaCl and silicate grains derived from the flux balance condition. In Fig.~\ref{Potential} (i) and (iii) the calculation of $\phi_d(a)$ using Eq. (\ref{extsecurrent}) is shown. The equilibrium potential for grains with radii $1\text{nm} \lesssim a\lesssim0.1\mu\text{m}$ differs from grains with radii $a\gtrsim0.1\mu\text{m}$ and for NaCl may even be of different sign. In the curves (ii) and (iv) of Fig.~\ref{Potential} the equilibrium potential from a conventional approach to $J_s$ is plotted (Meyer-Vernet \citeyear{MeyerVernet}) using the Sternglass formula. This approach does not include the small particle effect.
In Fig. \ref{Potential} (bottom, left) we show the dependence of the charge-to-mass ratio $q/m$ on the grain radius.
The curve (i) of Fig.\ref{Potential} (bottom, left) shows the calculation of $q/m$ using Eq. (\ref{extsecurrent}). The charge-to-mass ratio obtained from a conventional approach to $J_s$ is plotted in the curve (ii) of Fig.\ref{Potential} (bottom, left). The results presented in the Fig. \ref{Potential} (bottom, right) emphasize that the two SEE models may lead to different signs of the charge.

In Fig.~\ref{BigPotential} we show the calculated potential for spherical grains of different radii $a$ as a function of radial distance from Jupiter, in the range $4R_J\leq r\leq8R_J$. Depending on grain size the resulting equilibrium potential $\Phi_d(a,r)$ changes its sign. 
The change in sign arises from a variation of plasma parameters and the grain size dependence comes from SEE, which in turn is most sensitive to the densities and temperatures of the cold and warm electrons. The potential $\Phi_d(a,r)$ varies from $-15\text{V}$ to $+7\text{V}$.
In an earlier model by \citet{Horanyi96} the parameters $\delta_m=1$ and $E_m=500\text{ eV}$ were used, and the Sternglass formula for SEE was employed. In that framework a surface potential of $\Phi_d(r)\approx-30\text{ V}$ was obtained for the cold plasma torus ($4R_J\leq r\leq6R_J$) and + 3V elsewhere. In this orbital range, the plasma density peaks close to the orbit of Io ($r\sim6R_J$) with $n_{e}^{\text{cold}}\approx3700\text{ cm}^{-3}$ and plasma temperature has a strong minimum $kT_{e}^{\text{cold}}\approx1\text{ eV}$ at $r\sim5R_J$ and rises to $kT_{e}^{\text{cold}}\approx55\text{ eV}$ at $r\sim7.9R_J$. The temperature of ions and warm electrons (for $4R_J\leq r\leq8R_J$) are $kT_{i}^{\text{cold}}\approx30\text{ keV}$ and $kT_{e}^{\text{warm}}\approx1\text{ keV}$ respectively.

\subsection{Charging time}
\begin{figure}[ht]
\includegraphics[scale=0.3]{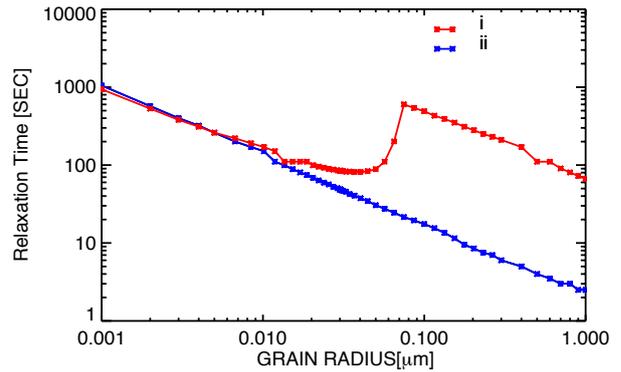}
\includegraphics[scale=0.3]{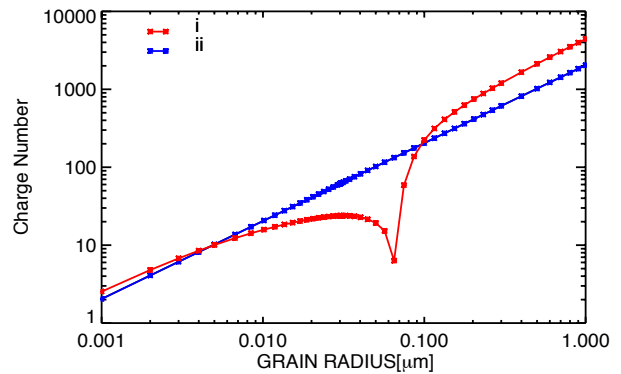}
\caption{(Color online). (Top) Dependence of the relaxation time $\tau_d$ for charge deviations on grain's radius $a$. The results presented for NaCl grains immersed in typical plasma conditions expected at $r\sim5.7R_J$ distance in the Jupiter system (see text). The result (i) is obtained for the generalized $J_s$ model (\ref{extsecurrent}) and compared to results (ii) obtained using the Sternglass approximation for SEE; (Bottom) Dependence of the charge number $\left|\left<N\right>\right|$ on grain's radius $a$ with the same parameters and notation as for Fig.~\ref{RelaxTimeChargeNum} (top).}
\label{RelaxTimeChargeNum}
\end{figure}
In Fig. \ref{RelaxTimeChargeNum} (top) we present results for the charging times $\tau_d(a)$ of NaCl grain of different radii.
Because the fastest charging occurs for high plasma densities in a close vicinity to Io, the results in Fig. \ref{RelaxTimeChargeNum} (top) could be 
considered as a reasonable estimate of the lower limit for $\tau_d(a)$ of NaCl grains in the Jupiter plasma environment (in the orbital range $[4R_J, 8R_J]$).
In the curve (i) of Fig. \ref{RelaxTimeChargeNum} (top) the calculation of $\tau_d(a)$ using Eq. (\ref{extsecurrent}) is shown.
In the curve (ii) of Fig. \ref{RelaxTimeChargeNum} (top) the charge relaxation time from a conventional approach to $J_s$ is plotted 
(Meyer-Vernet \citeyear{MeyerVernet}). From the curve (i) of Fig. \ref{RelaxTimeChargeNum}  (top) we see that the small particle effect modifies the well-known size dependence $\tau_d\sim a^{-1}$ relevant for the case described by the curve (ii). Thus, the different SEE models lead to different charging times.
\subsection{Estimate of charge fluctuations}
Because of the statistical nature of the collection currents, the calculated charge on the NaCl grain is not fixed, but it can fluctuate around its equilibrium value $q$ (Ciu \& Goree \citeyear{Ciu1994}, Meier et al. \citeyear{Meier2014}, Meier et al., \citeyear{Meier2015}). The charge fluctuation may also be in response to varying conditions in the Jupiter plasma environment.
It is shown that the quantized grain charging corresponds to a Poisson process (Khrapak et al., \citeyear{Khrapak}, Hsu et al., \citeyear{Hsu2011}) and the resulting distribution of charge fluctuations has the shape of a Gaussian with $\left<q_d\right>\sim q$ (Yaroshenko \& L\"uhr \citeyear{Yaroshenko}). Based on Poisson statistics it was obtained (Ciu \& Goree \citeyear{Ciu1994}, Khrapak et al., \citeyear{Khrapak}) that the fractional charge fluctuation varies as $\Delta q_d/\left<q_d\right> \sim \left|\left<N\right>\right|^{-1/2}$, where $\left<N\right>=\left<q_d/e\right>$ is the charge number obtained from the continuous charging model. 

In the curve (i) of Fig. \ref{RelaxTimeChargeNum} (bottom) the charge number $\left<N\right>$ from a conventional approach to $J_s$ is plotted. In the curve (ii) of Fig. \ref{RelaxTimeChargeNum} (bottom) we show results for the Sternglass model for SEE. 
From curve (ii) of Fig. \ref{RelaxTimeChargeNum} (bottom) the $\left|\left<N\right>\right|^{-1/2}$ scaling means that smaller nano-sized NaCl grains may have larger fractional charge fluctuation levels. However, from the curve (i) of Fig. \ref{RelaxTimeChargeNum} (bottom) it can be seen that for our generalized $J_s$ model (\ref{extsecurrent}) the NaCl grains with radii close to $a \sim3$ nm and $a \sim60$ nm have comparable values of the charge number $\left|\left<N\right>\right|$. Thus, the small particle effect modifies the dependence of the fractional charge fluctuations $\Delta q_d/\left<q_d\right>$ on the grain size.

\subsection{Field emission and grain charge limitation}
Even a relatively small charge on a grain of small size may induce a large surface electric field, which modifies significantly the potential barrier at the surface (Mendis \& Axford  \citeyear{MendisAxford}). So, the high-field processes of electron or ion field emissions, in general, have to be taken into account.

The electric field may be high enough to eject electrons from the grain surface. This may limit the negative charge of the grain. The electron field emission limit of surface electric field ($\left|E\right|=\left|\phi_d\right|/a$) is about $10^{3} \text{ V}/\mu\text{m}$ (Gomer \citeyear{Gomer}). It corresponds to the limiting potential $\phi_{ef}\approx -1\text{V }(a/1\text{nm})$ which is equivalent to a limiting grain charge 
$q_{fe}\sim -0.7 e (a/1\text{ nm})^2$. For instance, for grains with $a<1.7\text{ nm}$ we have $\left|\left<N\right>\right| < 2$, and the grain charge is limited to one single electron. For grains with $a<0.1 \mu\text{m}$ and $a<1 \mu\text{m}$ we get $\left|\left<N\right>\right|| < 7\cdot 10^{3}$ and 
$\left|\left<N\right>\right| < 7\cdot 10^{5}$. Thus, as it can be seen from the results in Fig. \ref{BigPotential} and, in particular, from the curve (i) of Fig. \ref{RelaxTimeChargeNum} (bottom) the negatively charged NaCl grains of radii $a\gtrsim60\text{ nm}$ have an equilibrium charge that is much lower than the limiting charge. So, the field electron emission does not show an effect for the NaCl grains. The field ion emission which limits the positive charging, requires a much higher field than the field electron emission $\sim 10^{4}\text{ V}/\mu\text{m}$ (Good \& M\"uller \citeyear{GoodM}) which corresponds to the limiting potential $\phi_{if}\approx +10\text{V }(a/1\text{nm})$. Accordingly, for positively charged grains with $a\lesssim60\text{ nm}$ we have $\left|\left<N\right>\right| \lesssim 2.5\cdot 10^{4}$. Thus, from the results in Fig. \ref{BigPotential} follows that the ion field emission does not show an effect for positively charged NaCl grains also.

\section{Conclusions}
In this study, we have revised the treatment of the charging of small grains immersed in an isotropic flux of electrons. 
For this purpose, we obtained a new model for the secondary electron current $J_{s}$ due to a generalization of the SEE model valid for small grains immersed in an isotropic flux of electrons. Furthermore, we qualitatively and quantitatively show how the results differ from widely used conventional models for the SEE yield. 
Our generalization of the charging model is important for dust grain sizes typically found in circumplanetary environments. 

Employing plasma conditions for the Jovian system, we find that for small ($a\leq1\mu\text{m}$) particles the grain potential $\phi_d$ depends on grain size while this is independent of size for big ($a\geq1\mu\text{m}$) particles. 
This effect may have important consequences for circumplanetary dust:
\begin{itemize}
\item[--] two grains (or parts of a more complex body), with the same composition and shape in the same environment and with the same history, can have different charges, depending on their sizes, in extreme cases they can have charges of opposite sign;
\item[--] the grain potential may significantly differ depending on grain size;
\item[--] the new SEE model leads to estimates of charging times that are qualitatively different from similar estimates derived from the traditional SEE model.
\item[--] nano-sized and sub-micron sized grains may have comparable fractional charge fluctuations;
\item[--] field emission does not limit the grain charge.
\end{itemize}

We also show that the charging in the Jupiter environment is particularly sensitive to the dust particle's material properties. Because the dynamics of grains in the range between 10 nm to 100 nm is dominated by the electromagnetic field, the precise value of the grain charge has a strong effect for particle transport and ejection from the Jovian system. 
We used our generalization of the charging model to estimate the surface potential of dust grains in the Jupiter environment but the results have a wider range of applicability and can be relevant for many charging processes where secondary electron currents are involved.
\begin{acknowledgements}
      This work was supported by German
      \emph{Deut\-sche For\-schungs\-ge\-mein\-schaft, DFG\/} project
      through Schwerpunktprogramm 1488 "Planet Mag" (Grant SCHM 1642/2--1) and by European Space Agency (ESA), funding project Jovian 
      Meteoroid Environment Model (JMEM, contract number: 4000107249/12/NL/AF). 
\end{acknowledgements}

\end{document}